\documentclass[conference]{IEEEtran}
\ifCLASSINFOpdf
 
\else
 \fi
\usepackage{graphicx}
\usepackage{amsmath}
\usepackage{amsbsy}
\usepackage{amsthm}

\usepackage{bbm}
\usepackage{bm}
\usepackage{epsfig}
\usepackage{epstopdf}
\usepackage{dsfont}
\usepackage{amsfonts}
\usepackage{amssymb}
\usepackage{calligra}

\def\EQ#1{\begin{eqnarray}#1\end{eqnarray}}
\newcommand{\djj}{d\kern-0.4em\char"16\kern-0.1em}

\newcounter{lem}

\newtheorem{prop}{Proposition}\def\PRO{\begin{prop}}\def\ORP{\end{prop}}
\newtheorem{coro}{Corollary}\def\COR{\begin{coro}}\def\ROC{\end{coro}}
\newtheorem{theo}{Theorem}\def\TH{\begin{theo}}\def\HT{\end{theo}}
\def\TH{\begin{theo}}\def\HT{\end{theo}}
\newtheorem{defi}[prop]{Definition}\def\DE{\begin{defi}}\def\ED{\end{defi}}
\newtheorem{lemme}[lem]{Lemma}\def\LE{\begin{lemme}}\def\EL{\end{lemme}}

\def\ket#1{\left| #1 \right\rangle}

\begin{document}
%
\title{Advances in Quantum Reinforcement Learning}

\author{\IEEEauthorblockN{Vedran Dunjko$^\ast$}
\IEEEauthorblockA{Institute for Theoretical Physics,\\
University of Innsbruck,\\
Innsbruck 6020, Austria\\
\emph{now at:} \\
Max Planck Institute of Quantum Optics\\
Garching 85748, Germany\\
Email: vedran.dunjko@mpq.mpg.de}
\and
\IEEEauthorblockN{Jacob M. Taylor}
\IEEEauthorblockA{Joint Quantum Institute
\&
NIST,\\
College Park, Maryland 20742, USA\\ 
Email:  jmtaylor@umd.edu}
\and
\IEEEauthorblockN{Hans J. Briegel}
\IEEEauthorblockA{Institute for Theoretical Physics,\\
University of Innsbruck\\
Innsbruck 6020, Austria, and \\
Department of Philosophy,\\
University of Konstanz,\\
Konstanz 78457, Germany\\
Email: hans.briegel@uibk.ac.at}}
\maketitle
\begin{abstract}
In recent times, there has been much interest in quantum enhancements of machine learning, specifically in the context of data mining and analysis. Reinforcement learning, an interactive form of learning, is, in turn, vital in artificial intelligence-type applications. Also in this case, quantum mechanics was shown to be useful, in certain instances. Here, we elucidate these results, and show that quantum enhancements can be achieved in a new setting: the setting of learning models which learn how to improve themselves -- that is, those that meta-learn. While not all learning models meta-learn, all non-trivial models have the potential of being ``lifted'', enhanced, to meta-learning models. Our results show that also such models can be quantum-enhanced to make even better learners. In parallel, we address one of the bottlenecks of current quantum reinforcement learning approaches: the need for so-called oracularized variants of task environments. Here we elaborate on a method which realizes these variants, with minimal changes in the setting, and with no corruption of the operative specification of the environments. This result may be important in near-term experimental demonstrations of quantum reinforcement learning.
\end{abstract}

\IEEEpeerreviewmaketitle

\section{Introduction}

\footnote{\copyright 2017 IEEE. Personal use of this material is permitted. Permission from IEEE must be obtained for all other uses, in any current or future media, including reprinting/republishing this material for advertising or promotional purposes, creating new collective works, for resale or redistribution to servers or lists, or reuse of any copyrighted component of this work in other works.
Full citation:\\
V. Dunjko, J. M. Taylor, H. J. Briegel, Advances in quantum reinforcement learning, IEEE SMC, Banff, AB, 2017, pp. 282-287. doi: 10.1109/SMC.2017.8122616 (2017).  }The interest in the potential of quantum enhancements in aspects of machine learning (ML) and artificial intelligence (AI) has been on the rise in recent times.
For example, quantum algorithms which can be used for efficient classification and clustering \cite{2014_Wittek, 2013_Lloyd, 2013_Aimeur, 2014_Rebentrost}, have been proposed. 
Such classification and clustering tasks, more generally referred to as supervised and unsupervised learning,  form two out of three canonical branches of ML. These learning modes are complemented by reinforcement learning (RL)  \cite{SuttonBarto98}.  RL is central to some of the most celebrated recent successes in ML (e.g. a computer beating a human champion in the game of Go \cite{2016_Silver}), and is indispensable in the context of AI \cite{2003_Russel}. Nonetheless, it has received significantly less attention from the quantum information processing (QIP) community, in comparison to the (un)supervised learning settings. 
Few works do exist: for instance in  \cite{2005_Dong} it was suggested QIP may yield new ways to evaluate policies, and in \cite{2014_Paparo} it was proven that the computational complexity of a particular model, Projective Simulation \cite{2012_Briegel}, can be quadratically reduced. Arguably, one of the key reasons why RL has been less explored stems from important differences to (un)supervised learning, which significantly influence how, and to what extent, QIP techniques can be applied. 
In particular, standard (un)supervised learning methods deal with a stationary dataset -- a sample set from an (empirical) joint distribution $P(X\!\!=\!x,Y\!\!=\!\!y)$, giving data points $(\{x_i\})$ and corresponding classes ($\{y_i\}$) in the supervised setting, or just the bare points sampled from $P(X\!\!=\!\!x)$ in the unsupervised case. For convenience, we will abbreviate (un)supervised learning with DL (for data-learning).
 In contrast, RL is about interactive, reward-driven learning, involving a task environment. The most prominent feature of RL is that the learner, typically called the \emph{learning agent}, influences the state of the environment -- thereby changing the distribution of what it perceives. This distinction makes the RL model particularly well-suited for AI-type applications (rather than data analysis). The same distinction causes additional technical obstacles in defining quantum analogs, or enhancements, for RL, relative to DL. To clarify, since in DL we are dealing with a stationary database, we can think of it as being encoded in a memory where data points are accessible via an address (such as RAM or hard and flash drives). It will be convenient to think of such memory as a process, which, given an address or index, outputs the corresponding data point: $i \stackrel{Mem}{\rightarrow} x_i.$ Such a memory can easily be represented as a reversible process, and consequently as a quantum operation, a unitary map, performing $\ket{i}\ket{0} \stackrel{qMem}{\rightarrow}\ket{i}\ket{x_i}.$ In \cite{2008_Giovannetti} it was shown how such a unitary can be implemented efficiently. Having access to such an object often allows for the efficient encoding of data into amplitudes, which is at the basis of many exponential quantum improvements \cite{2014_Rebentrost, 2016_Schuld}, or directly with the purpose of quadratic improvements \cite{2013_Aimeur}. The memory can be thought of as a black-box oracle and quantum-parallel access allows for query-complexity improvements, a topic which has been exhaustively explored from the early days of quantum computation. Note, in such settings, overall improvements are only attainable assuming such a memory has already been ``loaded'' with the data-set off-line, barring few exceptions \cite{2016_Lloyd_NC}. 
In the RL case, however, the environment cannot be modelled as a stationary memory, as the environmental responses depend on the interaction history. In particular, they depend on the actions of the agent. The environment is thus a process with a memory of its own, which cannot be accessed by the agent -- a memory with memory, so to speak. The general framework for quantum (RL) agents which interact with quantum environments was provided first in \cite{2016_DunjkoPRL} and in the technical preprint \cite{2015_Dunjko} parts of which we summarize and expand upon here. 

In the most general setting, both the learner and the environment are maps with memory (often called quantum combs \cite{2008_Chiribella}), and classical RL corresponds to the case where the maps are themselves classical. Any improvement beyond that of computational complexity requires that the environment allows some kind of quantum coherent access. This is an intuitive observation (and formally proven in \cite{2015_Dunjko}), in the same sense a classical phone book cannot be ``Groverized'' -- it has to be first be mapped to a system which can act on superpositions.
Thus this claim also applies for the case of DL, where the suitable quantum analog is straight-forward -- it is the map $qMem,$ see Fig \ref{fig2}. For the RL case, the situation is much less clear. In \cite{2016_DunjkoPRL} it was shown that a quantum analog, which allows for provable learning improvements, does exist. In this work, we will expand on these results two-fold. We will show how such a map can be black-box constructed given any specification of a task environment, and consider cases when such a construction is possible. Furthermore, we will identify a new family of improvements, based on so-called meta-learning, which are possible given such an oracular construction. To do so, we first present the broad framework, outlined in \cite{2016_DunjkoPRL}, and the ideas of the key results we will need later.

\subsection{Classical and quantum agent-environment interaction}
The standard turn-based RL paradigm comprises the percept $(\mathcal{S})$ and action  $(\mathcal{A})$ sets which specify the possible outputs of the environment, and the agent, respectively. The agent and the environment interact by sequentially exchanging elements from the percept/action sets. A realized interaction up to time step $t$, between the agent and the environment, that is a sequence $h_{t} = (s_1, a_2, s_3,s_4, \ldots, s_{t-1},a_{t}), s_i\in \mathcal{S}, a_j \in \mathcal{A}$ of alternating percepts and actions is called \emph{the $t-$step history} of interaction. At the $t^{th}$ time-step, and given the elapsed history $h_{t-1},$ 
the behavior of the agent at step $t$ is given by the map $M_{A}^{h_{t-1}}(s\in\mathcal{S} ) \in distr(\mathcal{A}),$

where $distr(\mathcal{X})$ denotes the set of probability distributions over the set $\mathcal{X}.$ The realized agent's action, given history $h_{t-1}$, is sampled from the distribution $M_{A}^{h_{t-1}}(s\in\mathcal{S} )$. The environment is specified analogously.

The notion of a \emph{reward} $\lambda \in \Lambda$ in RL (specifying whether a performed action, or a sequences thereof were `correct') can be w.l.g. subsumed into the percept space.
All the standard figures of merit regarding the learning performance of an agent are functions of the realized history. They are often convex-linear, enabling statements about average performance. Thus, the interaction history is the fundamental object in RL, and must be maintained in the quantum setting.

In an extension of the above framework to a quantum setting, the percepts/actions are promoted to orthogonal states (\textit{percept}/\textit{action} states) of the percept/action Hilbert spaces $\mathcal{H}_\mathcal{S} = \textup{span}\{\ket{s_i}\}_i $ and $\mathcal{H}_\mathcal{A}=\textup{span}\{\ket{a_i}\}_i .$ The agent, and the environment, contain internal memory:  finite (but arbitrarily sized) \textit{internal registers} $R_A$ and $R_E$
which can store histories, with Hilbert spaces of the form $\mathcal{H}_\mathcal{A} \otimes\mathcal{H}_\mathcal{S} \otimes \mathcal{H}_\mathcal{A}  \cdots$. 
We jointly call the percept/action states, their probabilistic mixtures, and sequences thereof, \emph{classical states}.
The interaction is modelled via a common \emph{communication register} $R_C$, with associated Hilbert space $\mathcal{H}_C~=~\{ \ket{x} | x\in \mathcal{S} \cup \mathcal{A} \}$ sufficient to represent both actions and percepts whose spaces are mutually orthogonal\footnote{A more general definition of an interaction, where in the spirit of robotics and embodied cognitive sciences we separate the interfaces of the agent and the environment, is provided in \cite{2015_Dunjko}.}.
The agent (environment) is fully characterized by sequences of completely positive trace preserving (CPTP) maps  $\{\mathcal{M}^{A}_i \}_i$ ($\{\mathcal{M}^{E}_i \}_i$) acting on the concatenated registers $R_AR_{C}$ ($R_{C}R_{E}$). We assume the three registers are pre-set in a fiducial classical product state.
An agent-environment interaction is then defined by the maps which are sequentially applied.

\begin{figure}[!h]
\underline{{\small \textsf{RL:}}}\\
\includegraphics[width=0.47\textwidth, trim=0cm 12.9cm 0cm 12.4cm,clip=true]{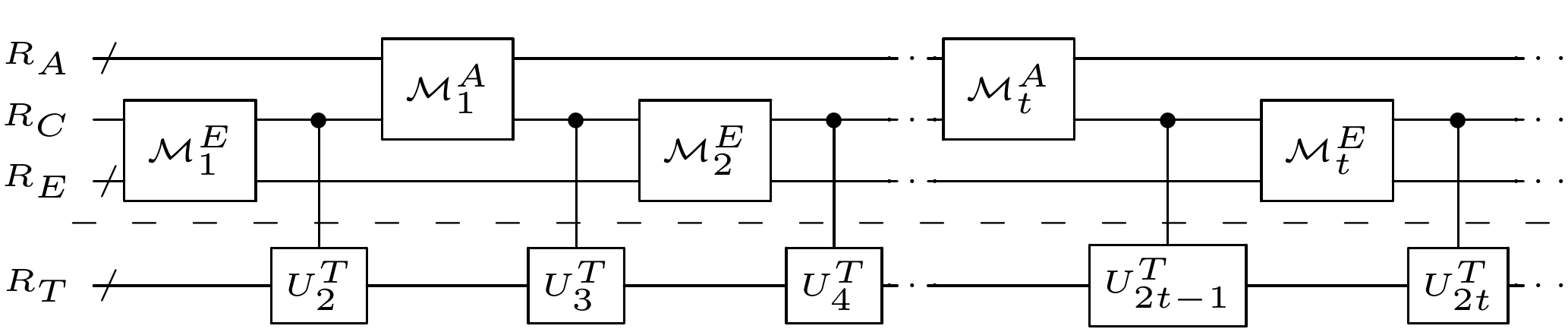}\\
\vspace{-1cm}\underline{{\small \textsf{DL:}}}\\
\includegraphics[width=0.47\textwidth, trim=0cm 12.9cm 0cm 12.4cm,clip=true]{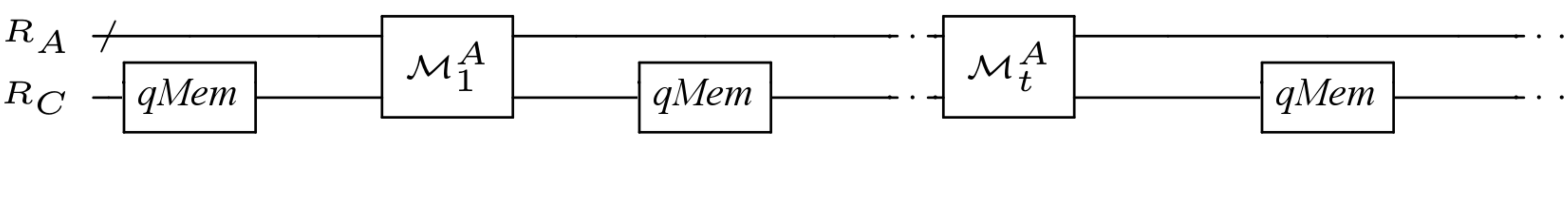}\\
\vspace{1cm}
\caption{\label{fig2}
RL: Tested agent-environment interaction suitable for reinforcement learning. In general, each map of the tester $U^T_k$ acts on a fresh subsystem of the register $R_T$, which is not under the control of the agent, nor of the environment. The crossed wires represent multiple systems.
DL: The simpler setting of standard quantum machine learning, where the environmental map is without internal memory, presented in the same framework. See main text for further detail.
}
\end{figure}

\emph{Classical agents} (environments) are those whose maps do not generate non-classical states (superpositions of classical states), given classical states of the internal and the communication registers. More generally, an agent and environment have a \emph{classical interaction} if the joint state of registers $R_AR_{C}R_E$ is separable w.r.t. the three partitions, and the state of $R_{C},$ post-selected on any outcome of any separable measurement of $R_AR_E$ is a classical state,  at every stage of the interaction\footnote{Classical interaction still allows that the internal information processing of the agent and environment includes non-classical states, e.g. an internal quantum computer. However, neither the agent or environment are allowed to output non-classical states, or to be entangled to the communication register $R_C$.}. To maintain a robust notion of history, we introduce \emph{testers,}  systems which monitor the interaction, and record it in its own memory $R_T$. Testers we consider are sequences of controlled maps of the form 
 \EQ{
U^T_k \left( \ket{x}_{R_C} \otimes \ket{\psi}_{R_T} \right) = \ket{x}_{R_C} \otimes U^{x}_k\ket{\psi}_{R_T} \label{tester}\nonumber
} 
where $x \in \mathcal{S} \cup \mathcal{A},$ and $\{ U^{x}_k\}_x$ are unitary maps acting on the register $R_T$, for all steps $k$. A tested interaction is shown in Fig. \ref{fig2}.
If all the maps of the tester copy\footnote{By `copy' we mean the map $\ket{x}\ket{\epsilon}\rightarrow\ket{x}\ket{x}\ \forall x\in\mathcal{S} \cup \mathcal{A}$, for some fixed state $\ket{\epsilon}$.} the classical states we call it a \emph{classical tester}. A \emph{sporadic classical tester} allows periods of untested interaction (i.e. some maps are identities).

\subsection{Quantum improvements in learning}
Improvements have been demonstrated for a few types of task environments, including epoch-type environments, where the state of the environment is periodically re-set. For example, once a chess game finishes, the board is re-set for the next game; similarly, in navigation tasks, e.g. maze navigation, once the agent walker) has found the exit (goal), it is brought back to the initial position, and the task is iterated.
Our overall approach is inspired by contrasting environments to oracles which appear in many quantum algorithms. 
Standard environments do not match the specification of oracles. For instance, the actions of the agent are lost to the environment, and not returned. Nonetheless, for epoch-type environments $E$, we can define \emph{oracular instatiations} $E_q$ of the classically specified environment $E$, which are unitary. With access to the instatiations, the agent can execute amplitude amplification \cite{2000_Brassard} to obtain an example of a rewarding sequence of actions more efficiently.

Environments $E$, which can be accessed both in their classical $(E)$ or oracularized instantiations $(E_q),$ we call \emph{controllable environments}.

The power to identify winning action sequences alone says little about the learning capacities of the agent. However, an \emph{exploration} stage, i.e. searching, must precede an \emph{exploitation} stage, and the correct interplay of these two phases is a well-studied problem in RL \cite{SuttonBarto98,2003_Russel}\footnote{A related interplay occurs in optimization problems, where a local minimum is often rapidly found, and can be used. However, this opens the possibility that we are missing out on a global minimum.}. This strongly suggest that (significantly) faster search should be useful for learning.   
To formalize this intuition we define \emph{luck-favoring settings}.
Briefly, a learning model $A$ and environment $E$ are luck-favoring, if a \emph{lucky} agent (one which by chance alone finds many correct sequences of action during an early stage) outperforms an \emph{unlucky} agent (one which does not) in $ E$, after this early stage. The performance is measured relative to a chosen figure of merit $R$. Note that most benchmarking task environments are luck favoring, with standard learning models, relative to the usual figures of merit (e.g. finite-horizon average reward).
When we  combine luck-favoring settings, with the capacities of quantum agents to explore faster, and the notion of a sporadic classical tester, we obtain the following result (Theorem 1 in \cite{2016_DunjkoPRL} simplified):

\vspace{0.15cm}
\noindent\textbf{Main Theorem} \emph{(informal)}
\emph{Given a classical learning agent $A,$ and a controllable, classically specified epochal task environment $E$ such that $(A,E)$ are luck-favoring relative to some figure of merit $R$, there exists a quantum agent $A^q$ which outperforms $A$ in $R,$ relative to a chosen sporadic classical tester.}
\vspace{0.15cm}

The basic idea of the proof is to have the quantum agent $A^q$ to use quantum untested access to $E_q$ to obtain instances of winning sequences in quadratically reduced time. Given these, $A^q$ will, internally, and by using no interaction steps, `train' a simulation of $A$. Eventually, the simulation will get lucky (call this successfully trained simulation $A_{lucky}$).  $A^q$ then relinquishes control to $A_{lucky}$, and forwards percepts and actions between $A_{lucky}$ and the environment, under a classical tester.  Since the setting is luck-favoring, the main claim follows. Although amplitude amplification yields quadratic speed ups in search,
 in some learning settings this can lead to even an exponential improvement (relative to the task environment size) in learning agent performance. Such exponential separations occur when lucky instances occur exponentially infrequently (e.g. mazes with low connectivity). In this scenario, a quantum agent can find a successful instance with near unit probability while the classical agent still has an exponentially small chance of the same outcome, for a chosen time interval. Thus this can constitute a relevant exponential improvement, for time-limited games.

In previous work, the enhancements were proven by ``oracularizing'' task environments, to achieve an (abstract) mapping summarized with the following expression
\EQ{
\ket{a_1, \ldots, a_M} \stackrel{E^q_{\textup{\textit{oracle}}}}{\longrightarrow} (-1)^{\Lambda(a_1, \ldots, a_M)} \ket{a_1, \ldots, a_M} \label{E-quantum},
}
where $\Lambda(a_1, \ldots, a_M)$ is $1$ iff the sequence of actions $a_1, \ldots, a_M$ results in a reward. This map is then further refined in \cite{2016_DunjkoPRL} to achieve similar improvements for a broader class of settings, but the basic idea is the same. Next we show that a similar philosophy, but relying on other oraculizations, can be used for improvements of different flavours, and finally briefly address the conceptual cost of ``oraculization'' from a practical perspective.

\section{Quantum improvements beyond state exploration}
The basic idea behind quantum improvements described above can be summarized in a sentence. We use quantum access to an (not fully characterized) environment to learn properties which can help us optimize an otherwise classical agent. This leads to the following three-step schema for identifying improvements: a) identify an ``indirect property'' of interest; b) identify a suitable oraculization (and the corresponding class of environments which are compatible with the oraculization) using which the indirect property will be ascertained and c) prove improvements -- i.e. prove that a quantum agent can  win overall, although it sacrifices valuable learning steps to ascertain the indirect properties \cite{2016_DunjkoPRL}. 
To exemplify this on the result of the previous section, the relevant property is ``path leading to a reward'',  the oracle is the blocked oracle marking such paths, and the proof follows from the restriction on luck-favoring environments, and first-success hitting times.

\begin{figure}[!h]
\includegraphics[width=0.47\textwidth, trim=2cm 7cm 5cm 3cm,clip=true]{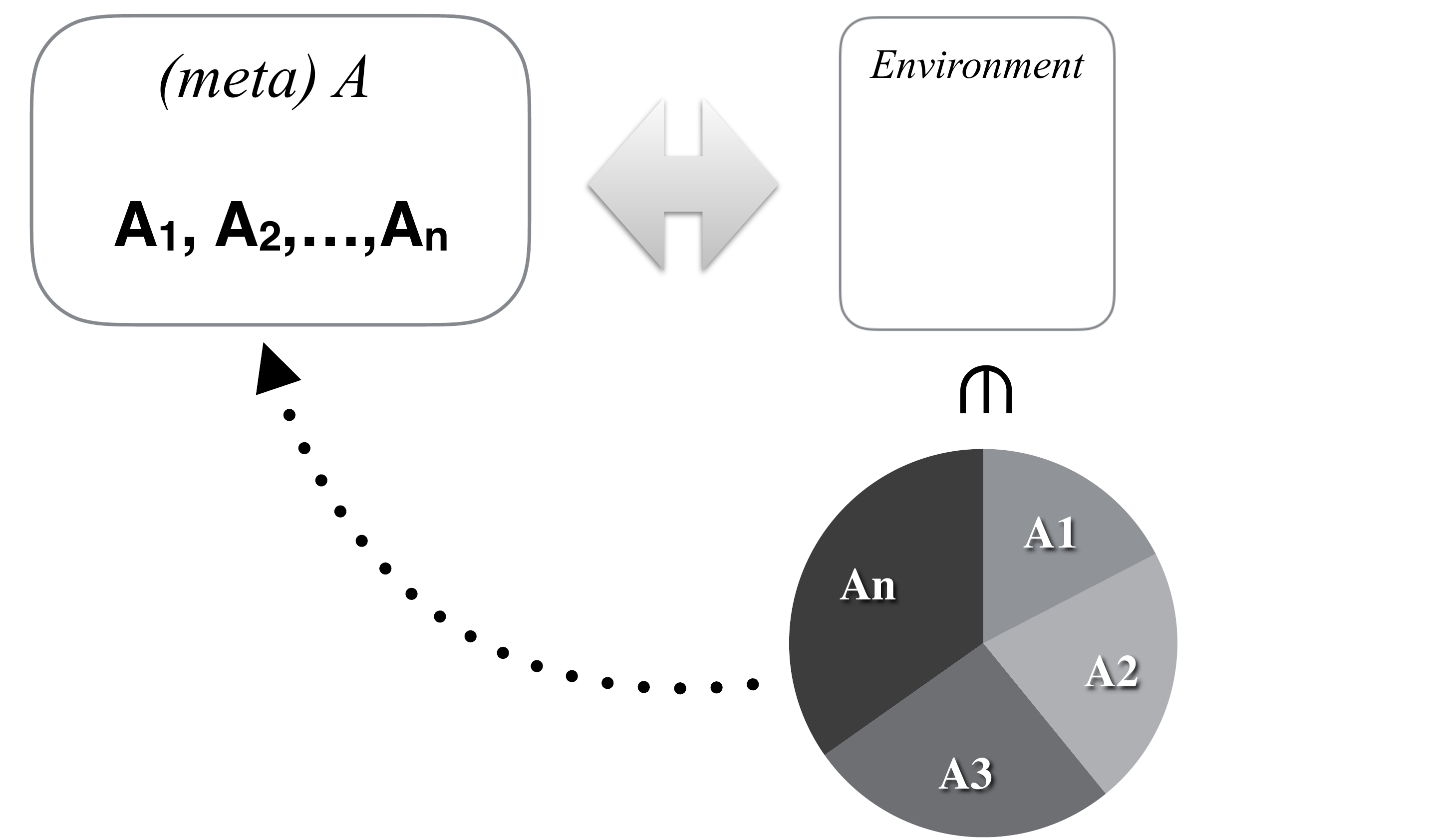}
\caption{\label{fig3}
Metalearning: The learning agent, over time, identifies the optimal configuration of its own metaparameters in order to optimize performance.
}
\end{figure}

A different way of obtaining improvements considers not just the environment, but also the agent at hand. 
Most learning models, both in the DL and RL case, come equipped with user-defined settings, called hyperparameters, or metaparameters, in the DL, and RL case, respectively. For example, if one uses neural networks for some task, one first needs to fix an architecture. In the case of RL, if one uses Q-Learning (or SARSA) \cite{2003_Russel}, they must choose the exploration/exploitation transition functions , that is, how quickly the model gets ``greedy''. The projective simulation (PS) model \cite{2012_Briegel} can have a few parameters including $\gamma$ (controlling the forgetting speed), $\eta$ (controlling reward propagation) and so on. The optimal learning settings of these parameters typically depends on the task environment at hand. This is, in fact, unavoidable because of No Free Lunch theorems \cite{1996_Wolpert}) which guarantee a universally optimal learner is not possible.
In practice, these parameters are set manually by the user. In general this should, and can, be automatized. In recent work, it was shown how these parameters can be learned in the context of the PS, but similar results have been obtained for other models as well (see \cite{2014_Melinkov}, and references therein). 
One way to think about this setting is to think of each configuration of parameter settings as fixing some particular learning agent/model $A_k$. On the other hand, the set of all considered environments is partitioned according to those for which a particular agent-configuration is optimal (out of the available set $\{A_k\}_k$). This is illustrated in Fig. \ref{fig3}.

Using the same oraculization approach as previously, we can now show how metalearning itself can be quantum-enhanced. The exact derivation of a possible approach is involved and long but the basic ideas are given next. First, note that, by taking a birds-eye perspective view of the scenario, any learning model with hyper/meta-parameters can be represented as a controlled unitary, controlled by a register specifying the values of the metaparameters (represented in the state $\ket{\mathbf{k}}_{m.m.p.},$ where $m.m.p$ stands for ``model metaparameters''). With $\mathbf{k}$ we denote any collection of such parameters. The ``model unitary'' is acting on the current memory of the agent (containing so-called model parameters, abbreviated $m.p.$), and the communication channel. As usual, we represent the environment in its purified picture, which simply means no information is lost. 
A learning agent without metalearning would then, through interaction with the environment, realize a joint agent-channel-environment state of the form
\EQ{
\ket{non-M.A.} = \ket{\mathbf{k}}_{m.m.p.}\ket{mem_A}_{m.p.}\ket{a}_C\ket{state}_E\nonumber,
}
where $mem_A$ is the memory of the agent, $a$ is the current action (assuming it is the agent's move), and $state$ is the purified state of the environment.
The expression above assumes both the agent and environment are described by deterministic programs. In the case either is randomized, the systems of the agent and environment would be entangled, with an extra system, specifying the random choices, but we ignore this for the moment.
Note that since everything is unitary, this overall state must determine the entire history of interaction. However $\mathbf{k}$ does not depend on it.
In order to metalearn the parameters, the standard approach is to have the agent monitor its own performance. Such a pre-metalearning agent would realize the joint state of the form
\EQ{
\ket{p.M.A.} = 
\ket{\mathbf{k}}_{m.m.p.}\ket{eval(\mathbf{k})}\ket{mem_A}_{m.p.}\ket{a}_C\ket{state}_E \nonumber,
}
where $eval(\mathbf{k})$ is some objective evaluation of the performance (the value depends on the environment, agent and $\mathbf{k}$), under settings $\mathbf{k}$, in $E$.

Now, in classical metalearning, a full metalearning agent must somehow optimize $eval(\mathbf{k})$ in $\mathbf{k},$ and this is is usually done by gradient descent. The idea here, naturally, is to use quantum processing to optimize this parameter. There are many proposals for quantum optimization (including recent \cite{2017_Rebentrost}), applicable in various cases, but many assume a special form of the function to be optimized. Since here we wish to maintain the presentation model-independent (albeit, at the cost of its efficiency) we approach this, for now, using the generic black-box, discrete optimization techniques such as \cite{1996_Durr}. Note, in doing so we have sacrificed the potential for exponential improvements.
Such approaches typically ``Groverize'' the process, and we proceed likewise here: The agent would first initialize its metaparameter register in a uniform superposition, so proportional to $ \sum_{\mathbf{k}}$, leading to the superposed state 
\EQ{
\ket{M.A_q.} \propto
\sum_{\mathbf{k}} \ket{\mathbf{k}}_{m.m.p.}\ket{eval(\mathbf{k})}\ket{mem_A}_{m.p.}\ket{a}_C\ket{state}_E \nonumber.
}
Note, so far we did not need to adjust the environment in any way, and if the agent were to now observe its evaluation register, it would collapse its history to match one particular choice of $\mathbf{k}$ (corresponding to $eval(\mathbf{k})$\footnote{The metaparameter register may still be in superposition if multiple choices lead to the same value.}). If it is lucky, or if this is repeated sufficiently many times, it would find the optimal choice, and this is eqivalent to a classical ``random guess'' agent. If the environment is epochal (as in \cite{2016_DunjkoPRL}), and controllable to the extent that we can reverse it, we can now amplitude amplify the history-branch corresponding to the best $\mathbf{k}.$\footnote{ 
Note, the capacity that the environment can be reversed implies either a deterministic environment, or that, in the register hijacking phase \cite{2015_Dunjko}, the agent can temporarily access the register purifying the random choices of the environment. However, it is interesting to consider purely deterministic environments as well, where the agent need only have access to its own randomness-inducing purifying register. This we can assume to be possible without loss of generality.}
  For this process to be feasible, it is not necessary to know the exact value of the optimal performance $eval(\mathbf{k}),$ but just a reasonable range. In this case, one can additionally use a binary search to pinpoint the optimal value while yielding only a logarithmic overhead, relative to the setting where the optimal value is known \cite{1996_Durr}).
The above method is optimal, and yields a strict quadratic improvement, when the mapping from metaparameters to performance has absolutely no structure.
Often, we are facing models at the opposite extreme where the parameter space has plenty of structure. For example, in \cite{2014_Melinkov}, the meta-parameter $\gamma$  -- the ``forgetting'' parameter, but the details are not relevant here -- of the PS model is, is optimized. The analysis of this work also suggests that, in many environments, the abstract mapping, which maps the parameter value to and averaged performance $[0,1] \stackrel{eval}{\rightarrow} \mathbbmss{R}^+,$ is an unimodal function -- it is increasing to some value $\gamma_{opt},$ and decreasing afterwards. This case is easy for classical learning, as one can perform a binary search. However, even here there are better options, and, in terms of expected time, probably the best option is using the quantum Bayesian adaptive learner \cite{2008_Ben-Or}, leading to a three-fold improvement (note, here improvements beyond multiplicative are not possible).
The intermediate regimes are more interesting, specifically when we deal with many continuous parameters. In this case one can, for instance, use ``quantum-improved'' gradient descent methods \cite{2005_Jordan}, or use ``completely quantum'' gradient descent techniques \cite{2017_Rebentrost}.
Once the learning model is fixed, it is likely that even more efficient methods become available, but this is beyond the scope of this paper.

\section{Cost of oraculization}

Most of the analyses we had done simply assume access to controllable environments, and consequently, to oracularized instantiations of the task environment. This is a reasonable assumption if the environments are constructed. If not, sometimes oracularized instantiations can nonetheless be effectively realized.
In particular, this is possible if we relax the model of interaction, and grant additional powers to the agent: \emph{register hijacking}, and \emph{register scavenging} \cite{2015_Dunjko}.
Hijacking pertains to the option that the agent has access to the register (memory) of the environment (but not to the specification of the environment it is actually learning), or to the purifying systems of the environment (which keep track of the history).
Here we show that these two options suffice implement the useful oraculizations in a (near) black-box fashion (i.e. without needing the specificities of the implementation).

For illustrative purposes, consider deterministic environments with one reward. For this particular case, we need to construct the map in Eq. \ref{E-quantum}.

We will give an explicit construction below, simplifying the steps given in  \cite{2015_Dunjko}, in which we assume the environment realizes a certain information transfer using specific gates -- in reality, whatever the environment does corresponds to \emph{some} choice of maps, which are isomorphic to the concrete choices we make below. In this  sense, our choices are without the loss of generality.
The steps of the construction are thus as follows. 

First using minimal register hijacking, the agent can, effectively,  realize the \emph{phase kick-back map}
\EQ{
\ket{a_1, \ldots, a_M}_A \ket{\epsilon, \ldots,\epsilon}_E \ket{\epsilon,\phi^-}_E \nonumber \\ \mapsto (-1)^R
 \ket{a_1, \ldots, a_M}_E \ket{s_2, \ldots, s_{M}}_A \ket{s_{M+1}, \phi^-}_A,\nonumber}
where the states $ \ket{\epsilon,\phi^-}$ and $\ket{s_{M+1}, \phi^-}$
 will be specified later. 

The map above is just a representation of $M$ steps of a standard interaction, where the agent committed to the action sequence $a_1, \ldots, a_M,$ to which the environment responded with  the percepts $\ket{s_2, \ldots, s_{M}} \ket{s_{M+1}, \phi^-}$ (hence, the agent no longer has access to the actions the environment stored/remembered.).
The register initialized in the state $\ket{\epsilon,\phi^-} $ is actually in the possession of the environment, and it should contain the memory slot where the environment will set the last percept, and the reward value. We can assume the reward value is flipped using a two-level unitary: Pauli-X $(\sigma_x)$, flipping between $\ket{0}$ to $\ket{1}.$ Here, the agent intervenes via hijacking, and imprints the state $\ket{\phi^-},$ which is the $-1$ eigenstate of $\sigma_x$. This will cause a phase kick-back.

Note that the map above is not yet the needed oracle as the action register is still entangled to the percept register.
Using a similar construction, the agent can also realize the raw map for the percepts (with no reward status):
\EQ{
\ket{a_1, \ldots, a_M}_A \ket{\epsilon, \ldots,\epsilon}_E \ket{\epsilon, \phi^+}_E \nonumber\\ \mapsto \ket{a_1, \ldots, a_M}_E \ket{s_2, \ldots, s_{M}}_A \ket{s_{M+1},\phi^+}_A.\nonumber  
}
The final ingredient of the construction is to note that any environment can be described using maps of  which are self-reversible.

Since the environment is classically specified, all the maps it realizes can be represented in a particular controlled form -- all classical computations also have this property. More precisely, the environments (can be assumed to) apply the map $E_t$ specified with
\EQ{
E_t \ket{a_1, \ldots, a_t} \otimes \ket{\epsilon}  = \ket{a_1, \ldots, a_t} \otimes E^{a_1, \ldots, a_t} \ket{\epsilon},
}
at each time-step $t$.
These maps $E^{a_1, \ldots, a_t}$  only rotate the fiducial (empty) state $\ket{\epsilon}$ to $\ket{s_{t+1} (a_1, \ldots, a_t)}$ (indicating also that $s_{t+1}$ depends on the previous actions). As clarified, the reward-setting unitaries act non-trivial only on a two-level subspace, but so are the maps $E^{a_1, \ldots, a_t}$, despite operating on an $|\mathcal{S}|$ - dimensional Hilbert space. But then, they can be chosen such that each $E^{a_1, \ldots, a_t}$ is Hermitian (and non-trivial only on a two-dimensional subspace), thus self-inverse.

The total process of oraculization is thus given by the following steps: a) The agent realizes the phase kick-back map, using hijacking; b) again, using hijacking, it implants the percept-containing systems (which it collected previously using scavenging) into the environment, at the correct memory location; c) the agent realizes the raw  map for percepts.
Given that the maps are all self-reversible (involutions),  in total  this realizes the oracular instantiation of $E_{oracle}.$
The total cost of interactions (number of interaction steps) with the environment we used was $2 M$, as two full games were played. 

While we naturally do not claim that the above process can be done with real macroscopic environments, it does show that there is no problem in principle, and that the definitional parts of the environment (i.e. what behavior is rewarded and so on), need not be touched to achieve oraculization. If the environment is say a computational environment of a future quantum network (which maintains the purity of all systems), the above is fully possible. It also becomes a possiblity in the case of nano-scale robots in future quantum experiments, where the environment is manifestly quantum and exquisitely  controled, as mentioned in \cite{2016_DunjkoPRL}. 
Critically, these processes are fully realizable to speed-up model-based learning, where the learner builds an internal representation of the environment, which can be fully quantum and controllable\footnote{Indeed, in this case it may be possible to build the oracular variant directly, however the described procedure shows how this can be done in a black-box fashion, i.e. without caring about the details of implementation.}.

\section{Conclusion}
The recent successes in classical AI and quantum machine learning point towards the possibilities of new, perhaps revolutionary, quantum-enhanced smart technologies. However, most such smart technologies, aside from performing extensive data analysis, must still learn from interaction.
This opens a ``quantization bottleneck'' which, for instance, prevents us from seriously discussing actual quantum artificial intelligence, (thus beyond clustering and classification). Progress in quantum reinforcement learning may help bridge that gap.
In this work, we have addressed the recent efforts in this field, and expanded the results for quantum improvements in learning quality, by showing how the process of learning how to learn -- metalearning -- can itself be quantum-enhanced. This yields quantum-enhanced learning agents which, generically, outperform their classical counterparts. As a side result, we have presented a succinct construction which shows how oracular instantiations of task environments, necessary for improvements, can be realized in a black-box fashion. This latter result may be relevant in future implementations of quantum reinforcement learning.

\section*{Acknowledgment}

VD and HJB acknowledge the support by the Austrian Science Fund (FWF) through the SFB FoQuS F 4012, and the Templeton World Charity Foundation grant TWCF0078/AB46.

\bibliographystyle{unsrt}

\begin{thebibliography}{10}
\expandafter\ifx\csname url\endcsname\relax
  \def\url#1{\texttt{#1}}\fi
\expandafter\ifx\csname urlprefix\endcsname\relax\def\urlprefix{URL }\fi
\providecommand{\bibinfo}[2]{#2}
\providecommand{\eprint}[2][]{\url{#2}}


  


\bibitem{2014_Wittek}
\bibinfo{author}{Wittek, P.} 
\newblock \emph{\bibinfo{title}{Quantum Machine Learning: What Quantum Computing Means to Data Mining}}
  (\bibinfo{publisher}{Academic Press}, \bibinfo{year}{2014}).



  
 \bibitem{2013_Lloyd}
\bibinfo{author}{Lloyd, S.}, \bibinfo{author}{Mohseni, M.} and
  \bibinfo{author}{Rebentrost, P.}
\newblock \bibinfo{title}{Quantum algorithms for supervised and unsupervised
  machine learning}.
\newblock \emph{\bibinfo{journal}{ArXiv:1307.0411}}  (\bibinfo{year}{2013}).

\bibitem{2013_Aimeur}
\bibinfo{author}{A{\"{i}}meur, E.}, \bibinfo{author}{Brassard, G.} and
  \bibinfo{author}{Gambs, S.}
\newblock \bibinfo{title}{Quantum speed-up for unsupervised learning}.
\newblock \emph{\bibinfo{journal}{Machine Learning}}
  \textbf{\bibinfo{volume}{90}}, \bibinfo{pages}{261--287}
  (\bibinfo{year}{2013}).

 \bibitem{2014_Rebentrost}
\bibinfo{author}{Rebentrost, P.}, \bibinfo{author}{Mohseni, M,} and \bibinfo{author}{Lloyd, S.} 
\newblock \bibinfo{title}{Quantum Support Vector Machine for Big Data Classification}.
\newblock \emph{\bibinfo{journal}{Phys. Rev. Lett. }} \textbf{\bibinfo{volume}{113}},
  \bibinfo{pages}{130503} (\bibinfo{year}{2014}).

\bibitem{SuttonBarto98}
\bibinfo{author}{Sutton, R.~S.} and \bibinfo{author}{Barto, A.~G.}
\newblock \emph{\bibinfo{title}{Reinforcement learning: An introduction}}
  (\bibinfo{publisher}{MIT Press, Cambridge Massachusetts},
  \bibinfo{year}{1998}), \bibinfo{edition}{first} edn.



\bibitem{2016_Silver}
\bibinfo{author}{Silver, D, \emph{et. al.}}
\newblock\bibinfo{title}{Mastering the game of Go with deep neural networks and tree search}
\newblock \emph{\bibinfo{journal}{Nature}} \textbf{\bibinfo{volume}{529}},
  \bibinfo{pages}{484--503} (\bibinfo{year}{2016}).
  

  \bibitem{2003_Russel}
\bibinfo{author}{Russel, S.~J.} and \bibinfo{author}{Norvig, P.}
\newblock \emph{\bibinfo{title}{Artificial intelligence - A modern approach}}
  (\bibinfo{publisher}{Prentice Hall}, \bibinfo{address}{New Jersey},
  \bibinfo{year}{2003}), \bibinfo{edition}{second edition} edn.






 \bibitem{2005_Dong}
\bibinfo{author}{Dong, D.}, \bibinfo{author}{Chunlin, C.} and
  \bibinfo{author}{Zonghai, C.}
\newblock \bibinfo{title}{Quantum Reinforcement Learning}.
\newblock \emph{\bibinfo{booktitle}{Advances in Natural Computation}}  
\newblock \bibinfo{series}{Lecture Notes in Computer Science}
\newblock \bibinfo{volume}{3611}
\newblock \bibinfo{pages}{686-689}
(\bibinfo{year}{2005}).




\bibitem{2014_Paparo}
\bibinfo{author}{Paparo, G. D.}, \bibinfo{author}{Dunjko, V.}, \bibinfo{author}{Makmal, A.}, \bibinfo{author}{Martin-Delgado, M. A.} and \bibinfo{author}{Briegel, H. J.}
\newblock \bibinfo{title}{Quantum speedup for active learning agents}.
\newblock \emph{\bibinfo{journal}{Phys. Rev. X}} \textbf{\bibinfo{volume}{4}},
  \bibinfo{pages}{031002} (\bibinfo{year}{2014}).

\bibitem{2012_Briegel}
\bibinfo{author}{Briegel, H.~J.} \& \bibinfo{author}{De~las Cuevas, G.}
\newblock \bibinfo{title}{Projective simulation for artificial intelligence}.
\newblock \emph{\bibinfo{journal}{Sci. Rep.}} \textbf{\bibinfo{volume}{2}}
  (\bibinfo{year}{2012}).
  
   \bibitem{2008_Giovannetti}
\bibinfo{author}{Giovannetti, V.}, \bibinfo{author}{Lloyd, S.} and \bibinfo{author}{Maccone, L.} 
\newblock \bibinfo{title}{Quantum Random Access Memory}.
\newblock \emph{\bibinfo{journal}{Phys. Rev. Lett. }} \textbf{\bibinfo{volume}{100}},
  \bibinfo{pages}{160501} (\bibinfo{year}{2008}).
  
 \bibitem{2016_DunjkoPRL}
\bibinfo{author}{Dunjko, V.}, \bibinfo{author}{Taylor, J. M.} \& \bibinfo{author}{Briegel, H. J.} 
\newblock \bibinfo{title}{Quantum-Enhanced Machine Learning}.
\newblock \emph{\bibinfo{journal}{Phys. Rev. Lett. }} \textbf{\bibinfo{volume}{117}},
  \bibinfo{pages}{130501} (\bibinfo{year}{2016}).

\bibitem{2015_Dunjko}
\bibinfo{author}{Dunjko, V.}, \bibinfo{author}{Taylor, J. M.} \&
  \bibinfo{author}{Briegel, H. J.}
\newblock \bibinfo{title}{Framework for learning agents in quantum environments}.
\newblock \emph{\bibinfo{journal}{arXiv:1507.08482}}  (\bibinfo{year}{2015}).

   \bibitem{2008_Chiribella}
\bibinfo{author}{Chiribella, G.}, \bibinfo{author}{D'Ariano, G. M.} \& \bibinfo{author}{Perinotti, P.} 
\newblock \bibinfo{title}{Quantum Circuit Architecture}.
\newblock \emph{\bibinfo{journal}{Phys. Rev. Lett. }} \textbf{\bibinfo{volume}{101}},
  \bibinfo{pages}{060401} (\bibinfo{year}{2008}).

\bibitem{2000_Brassard}
\bibinfo{author}{Brassard, G.}, \bibinfo{author}{Hoyer, P.},  \bibinfo{author}{Mosca, M.} and \bibinfo{author}{Tapp, A.}
\newblock \bibinfo{title}{Quantum Amplitude Amplification and Estimation}
\newblock \emph{\bibinfo{journal}{arXiv:quant-ph/0005055}}  (\bibinfo{year}{2000}).


 \bibitem{1996_Wolpert}
\bibinfo{author}{Wolpert, D.~H.}
\newblock \bibinfo{title}{The lack of a priori distinctions between learning
  algorithms}.
\newblock \emph{\bibinfo{journal}{Neural Comput.}}
  \textbf{\bibinfo{volume}{8}}, \bibinfo{pages}{1341--1390}
  (\bibinfo{year}{1996}).



\bibitem{2014_Melinkov}
 \bibinfo{author}{Makmal, A} , \bibinfo{author}{Melnikov, A. A.}, \bibinfo{Dunjko, V.} and \bibinfo{author}{Briegel, H. J.}
\newblock \bibinfo{title}{Meta-learning within Projective Simulation}.
\newblock \emph{\bibinfo{journal}{IEEE Access}}
\bibinfo{volume}{4}, 2110
  (\bibinfo{year}{2016}).
%
%


 \bibitem{1996_Durr}
\bibinfo{author}{Durr, C.}, \bibinfo{author}{Hoyer, P.} 
\newblock \bibinfo{title}{A Quantum Algorithm for Finding the Minimum}.
\newblock \emph{\bibinfo{journal}{arXiv:quant-ph/9607014}}  (\bibinfo{year}{1996}).

 


 \bibitem{2017_Rebentrost}
\bibinfo{author}{Rebentrost, P.}, \bibinfo{author}{Shuld, M.}, \bibinfo{author}{Petruccione, F.} and \bibinfo{author}{Lloyd, S.}
\newblock \bibinfo{title}{Quantum gradient descent and Newton's method for constrained polynomial optimization}.
\newblock \emph{\bibinfo{journal}{ arXiv:1612.01789v1}}  (\bibinfo{year}{2016}).

\bibitem{2005_Jordan}
\bibinfo{author}{Jordan, S. P.}
\newblock \bibinfo{title}{Fast Quantum Algorithm for Numerical Gradient Estimation}.
\newblock \emph{\bibinfo{journal}{Phys. Rev. Lett.}} \textbf{\bibinfo{volume}{95}},
  \bibinfo{pages}{050501} (\bibinfo{year}{2005}).


\bibitem{2016_Schuld}
\bibinfo{author}{Schuld, M.}, \bibinfo{author}{Sinayskiy, I.}, \bibinfo{author}{Petruccione, F.}
\newblock \bibinfo{title}{Prediction by linear regression on a quantum computer}.
\newblock \emph{\bibinfo{journal}{Phys. Rev. A}} \textbf{\bibinfo{volume}{94}},
  \bibinfo{pages}{022342} (\bibinfo{year}{2016}).
%
%
 \bibitem{2016_Lloyd_NC}
\bibinfo{author}{Lloyd, S.}, \bibinfo{author}{Garnerone, S.} \& \bibinfo{author}{Zanardi, P.} 
\newblock \bibinfo{title}{Quantum algorithms for topological and geometric analysis of data}.
\newblock \emph{\bibinfo{journal}{Nat. Comm.}} \textbf{\bibinfo{volume}{7}},
  \bibinfo{pages}{10138} (\bibinfo{year}{2016}).
%
%
%
%
%
%
%
%
%
%
%
%
%
%
%
%
%
%
%
%
%
%
%
%
%
%
\bibitem{2008_Ben-Or}
 \bibinfo{author}{Ben-Or, M.} and  \bibinfo{author}{Hassidim, A.},\newblock \bibinfo{title}{he Bayesian Learner is Optimal for Noisy Binary Search (and Pretty Good for Quantum As Well)}.
\newblock \emph{\bibinfo{journal}{Proceedings of the 2008 49th Annual IEEE Symposium on Foundations of Computer Science}} \textbf{\bibinfo{pages}{221-230}}
  (\bibinfo{year}{2008}).
%
%
%
%
%
%
%
%
%
%
%
%
%
%
%
%
%
%
%
%
%
%
%
%
%
%
%
%


\end{thebibliography}

\end{document}